# Deep-learning Segmentation of Small Volumes in CT images for Radiotherapy Treatment Planning

Jianxin Zhou[1], Kadishe Fejza[2], Massimiliano Salvatori[3], Daniele Della Latta[4], Gregory M. Hermann[2], Angela Di Fulvio[1]

1. Department of Nuclear, Plasma, and Radiological Engineering, University of Illinois at Urbana-Champaign, Urbana, IL 61801, United States.
2. Department of Radiation Oncology. OSF HealthCare. Peoria, IL 61637, United States.
3. IFC-National Research Council. Via Giuseppe Moruzzi, Pisa, Italy.
4. Terarecon Inc., 104309 Emperor Blvd, Suite 310 Durham, NC 27703, United States.

**Abstract**

**Purpose:** Our understanding of organs at risk is progressing to include physical small tissues such as coronary arteries and the radiosensitivities of many small organs and tissues are high. Therefore, the accurate segmentation of small volumes in external radiotherapy is crucial to protect them from over-irradiation. Moreover, with the development of the particle therapy and on-board imaging, the treatment becomes more accurate and precise. The purpose of this work is to optimize organ segmentation algorithms for small organs.
**Methods:** We used 50 three-dimensional (3-D) computed tomography (CT) head and neck images from StructSeg2019 challenge to develop a general-purpose V-Net model to segment 20 organs in the head and neck region. We applied specific strategies to improve the segmentation accuracy of the small volumes in this anatomical region, i.e., the lens of the eye. Then, we used 17 additional head images from OSF healthcare to validate the robustness of the V-Net model optimized for small-volume segmentation.
**Results:** With the study of the StructSeg2019 images, we found that the optimization of the image normalization range and classification threshold yielded a segmentation improvement of the lens of the eye of approximately 50%, compared to the use of the V-Net not optimized for small volumes. We used the optimized model to segment 17 images acquired using heterogeneous protocols. We obtained comparable Dice coefficient values for the clinical and StructSeg2019 images (0.61±0.07 and 0.58±0.10 for the left and right lens of the eye, respectively).
**Conclusions:** The segmentation results demonstrate that the optimized model has improved segmentation accuracy for small volumes and is robust in different clinical scenarios and the associated doses delivered to organs at risks are comparable to those obtained with manual segmentation. Therefore, the proposed approach can potentially aid the segmentation process and therefore improve the workflow of radiation therapy treatment planning.

**Keyword:**
Image segmentation, Deep learning algorithm, Radiation therapy treatment planning

## 1. Introduction

X-ray external radiation therapy heavily relies on treatment planning optimization to maximize the dose delivered to the gross target volume (GTV) while minimizing the dose to surrounding healthy tissues and ultimately reducing the risk of radiation-induced secondary cancers [1, 2, 3, 4]. Medical physicists acquire computed tomography (CT) images to outline the target volumes and organs at risk and use the labeled images to optimize radiation therapy treatment plans [5, 6, 7, 8, 9, 10]. This process is referred to as"segmentation." Manual segmentation is significantly time-consuming and strongly affected by intra- and inter-observer variations, even when segmenting according to the guidelines [11, 12]. Over the past three decades, many semi-automated segmentation methods have been developed. Most of them require image registration. Image registration is a process of transposing different sets of data into one coordinate system to align the same anatomical structures. Statistical models [13, 14] and "atlas" based methods [15, 16, 17, 18]are the two main segmentation approaches based on image registration. Statistical-based methods find common anatomical features in a reference data set and use them to develop the statistical model. Then the algorithm matches the target images to the model with informationabout the expected organ shape and appearance to conduct the segmentation. An "Atlas" is a library of pre-generated image data setswith reference manual segmentation. "Atlas" based methods register the target image with the "atlas" library to find the best subject match and propagate the reference segmentation to obtain the target segmentation result. The image registration algorithm significantly affects the accuracy and robustness of these semi-automatic segmentation methods. Therefore, more accurate and efficient segmentation methods, potentially fully automated and with better generalization capability, are needed.

Several deep-learning (DL)-based segmentation methods have been proposed to aid medical image segmentation. DL-based models feature, in most cases, encoder-decoder architectures, such as the fully convolutional network (FCN) [19] and U-Net [20]. The models derived from these architectures have achieved great success in recognizing and segmenting tumors and organs, and show promising segmentation accuracy on multi-organ areas, like the abdominal region [21, 22] and the head and neck region [23, 24, 25, 26]. However, the segmentation of small volumes remains challenging and researchers are developing different methods to improve their segmentation accuracy [27, 28]. The work presented in this paper focuses on the segmentation of small volumes in the head and neck region. The images that we used to develop the DL model are publicly-available three-dimensional (3-D) CT head and neck images [34]., which include manually segmented layers with 20 labeled soft-tissue organs, such as the brain stem, optic nerve, eyes, lens of the eyes, etc., listed in the next section. In these labeled images, the lens of the eye occupies the smallest volume. The lens of the eye is one of the most radiosensitive human tissues and the damage to cells covering its posterior surface can cause opacity and cloudiness in the lens. A threshold as low as 0.5 Gy for cataract formation has been described [29, 30, 31]. Therefore, accurate segmentation of the lens of the eye is needed to develop optimized treatment plans that minimize the dose to the eye during radiation therapy. The robust segmentation of small volumes is also a generally challenging problem. Methodologies that succeed in improving the segmentation of small volumes while maintaining a low computational complexity can be applied to larger volumes and provide better overall segmentation conformity.

We developed a DL model based on a V-Net [32], which can process 3-D data as a whole instead of processing them slice by slice to segment organs in the 3-D CT images. We also tested other architectures and found that the V-Net-based model can better define the organ boundaries under the complex environments of these 3-D image sets, compared to other neural networks, like U-Net [20]. In small volumes, a small localization bias may result in a severe segmentation error. In this work, we have demonstrated the feasibility of enhancing the V-Net model to segment 3-D CT images with three different methods to improve the segmentation of thee small volumes. The used methods are: 1) image preprocessing to enhance the image contrast; 2) optimization of the V-Net segmentation parameters specific to the small area; and 3) the application of an additional deep-learning algorithm (Mask R-CNN [33]) for the automated definition of the organ boundaries before segmentation. In addition to the publicly-available data sets that were used to develop the model, we also used a second data set provided by OSF HealthCare Department of Radiation Oncology under the University of Illinois College of Medicine at Peoria IRB 1 #00000688 to further validate the robustness of our model and strategies to improve the small volume segmentation using real-world clinical data.

**2. Computational Methods**

We have developed a V-Net algorithm to segment CT head and neck images, tested three approaches to improve the segmentation fidelity for small volumes, and used standard parameters of merit to quantify their performances.

*2.1. Data set*

We used two 3-D head and neck CT data sets in this paper. The first set encompassed 50 head and neck CT images manually segmented and labeled by experts within the framework of the StructSeg2019 Challenge. We used these images to develop the V-Net model and demonstrate image processing strategies to improve the segmentation of the lens of the eye. We used 35 data sets to train the model and five data sets to validate it to classify andsegment 20 different organs. Then, we used the remaining ten data sets as the test set to evaluate the model segmentation performance.The 20 soft-tissue organs are the brain stem, left eye, right eye, left lens of the eye, right lens of the eye, left optic nerve, right optic nerve, opticchiasma, left temporal lobes, right temporal lobes, pituitary gland, left parotid gland, right parotid gland, left inner ear, right inner ear, left mid ear, right mid ear, left temporomandibular joint, right temporomandibular joint, and spinal cord.

The second set of images included 17 external head and neck data sets from the OSF HealthCare. The images were acquired from patients who undergo CT simulation for planned radiation treatment to the head. The images included the entirety of the calvarium to the base of skull. The lens of the eye were manually segmented by a clinical research assistant and reviewed by a radiation oncologist. We used our model that was trained using the StructSeg2019 images to segment the lens of the eye in these images and further validate the model performance after applying the improvement strategies for small volumes.

The dimensions of images from both StructSeg2019 and OSF healthcare are 512×512×N. Each data set includes N number of 2-D 512×512 slices, with N ranging from 80 to 152. Fig. 1 (a) and (b) show a single 2-D slice of a head and neck image (StructSeg2019) and (c), and (d) show the head and neck region with minimum background surrounding it.

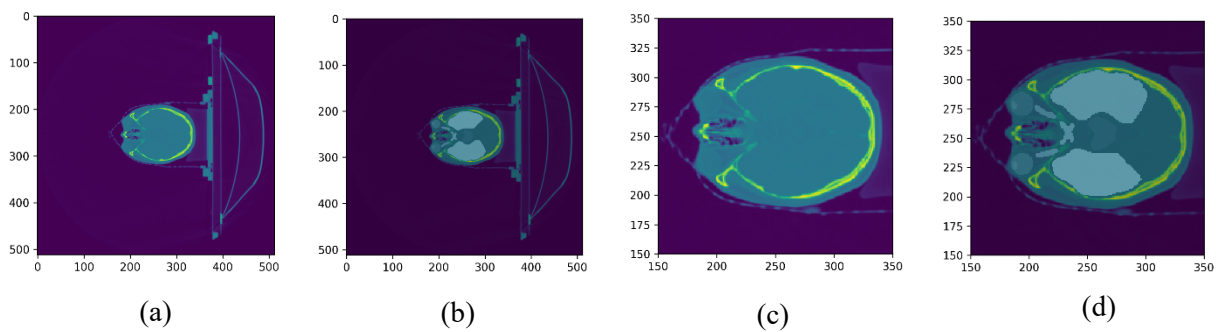

Figure 1: Example of a (a) non-labeled and (b) manually labeled slice selected from the original head and neck 3D-CT, (c) selected head and neck region, (d) with manually labeledorgans.

2.2. *V-Net model for segmentation*

The segmentation model that we have developed is based on a V-Net neural network. The V-Net is a fully convolutional neural network developed for medical image segmentation [32]. Fig. 2 shows the flowchart of the segmentation model.

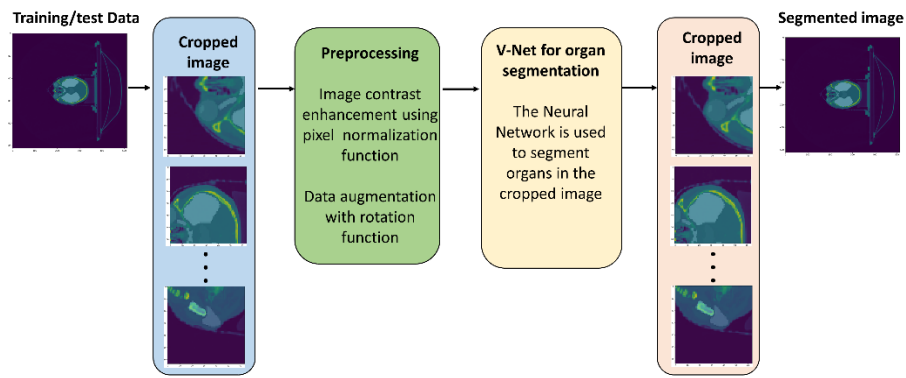

Figure 2: Flowchart of the algorithm used for training and testing the V-Net-based 3-D segmentation.

As described above, the dimensions of the 3-D images are 512×512×N, which is too large to be processed by the network at once. Therefore, the first step of the model flowchart is to crop the image and process a sub-volume of the image to reduce computational time, neglecting the large background area of the original image, shown in Fig. 1(a) (b). Based on the locations of the organs and tissues, we cropped the whole head and neck image into 12 portions. Each part could contain more than one organ. Fig. 3 (a) shows one of the 12 cropped image portions - the portion corresponding to the eye. We used this cropped image to trainthe model and segment the eye and the lens of the eye. The 50 StuctSeg2019 images that were used to train and test the model were anatomically similar. Therefore, a conservatively large eye cropping boundary was able to encompass the same anatomical target area around the eye for all images, as shown in Fig 3. However, the external test images from OSF healthcare were less homogenous in terms of patient anatomy, image resolutions, and image contrast. Therefore, a fixed cropping boundary could not be used. We developed an automated method to select the same areas around the eye for various images to overcome the dissimilarity of the OSF healthcare images. In the OSF data, we first used the segmentation model to locate the center of the eye, and then we took the center of the eye as origin to crop a 75 mm×75 mm×160 mm volume of the whole image in X-Y-Z directions, respectively. This method was able to select the same image portion of the eye for the OSF healthcare images and can be applied to any data set.

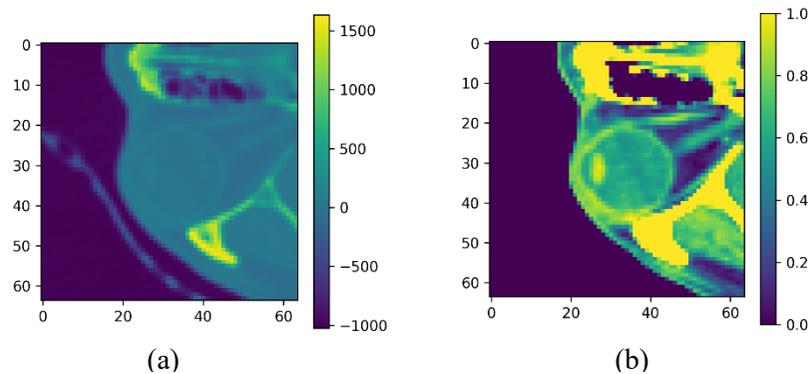

Figure 3: CT image subsection used as the input to the V-Net model, (a) eye region without locally optimized normalization, (b) eye region under 100 normalizationrange.

From Fig. 3 (a), we can observe that the original CT head and neck image has pixel values ranging from -1000 to +1500. However,the pixel values for the eye and the lens of the eye are in the one-hundred range. Therefore, with such pixel values, the eye and lens of the eye cannot be discriminated clearly, which is challenging for the V-Net classification and segmentation. In order to overcomethis issue, we performed a normalization step during the preprocessing, which limits the pixel values to a pre-determined window. InFig. 3(a), we applied a -100 to 100 normalization range to the image. Any pixel value larger than 100 was set to 100, any value smaller than -100 was set to -100, and values between -100 to 100 were normalized from 0 to 1. After this operation, the eye and lens of the eye are clearly visible in Fig. 3(b).

The following preprocessing step is data augmentation. The 35 3-D images available were not enough to train a sophisticated model with a high number of parameters. Therefore, we have applied a rotation function during the training process to augment the training set. For each image, we took the image center as the rotation center and rotated it along with three directions. The images were rotated from -25° to 25° by 3° steps. Each rotated image would be treated as a new image for our model. Therefore, we have created hundreds of images from the original 35 data sets to train the model.

After the cropping and preprocessing steps, we trained the model with each image portion and obtained the specific training weights to classify and segment organs inside this portion. Finally, we combined the segmentation results of each part to obtain the whole segmented image.

*2.3. V-Net architecture*

Fig. 4 shows our V-Net architecture. It is a 25-layer model, which consists of convolutional layers, max pooling layers, up sampling layers, dropout layers, and merging layers. Convolutional layers use filters to extract hidden features like edges, vertical lines, and horizontal lines of an image and store them in feature maps. This layer can reduce the data size and simplify regression and classification. The max pooling and up sampling layers adjust the dimensions of the feature maps to reduce computation time, and the merging layers combine two layers, giving our network the ability to recreate high-level features (like shapes) and also focus on other low-level features. The training process is based on this encoder and decoder neural structure and trains the model continuously until the loss graph approaches stability. We used the softmax [35] and cross-entropy [36] loss functions to calculate the loss in the training process, which is described in Section 2.4, and considered the graph to be stable once the relative error between subsequent samples was lower than 10%.

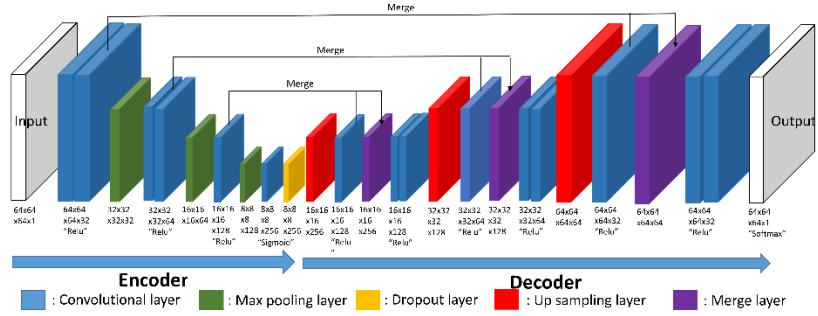

Figure 4: Architecture of the V-Net for 3-D head and neck image segmentation.

*2.4. Softmax activation and cross-entropy loss function*

The nature of our organ segmentation problem is a supervised machine learning problem for image classification, which classifiesthe organ type of each pixel of an input image. We first defined 20 target classes (organ types) and then provided a set of images, which are labeled with these organ types, and used them to train the neural network. When performing segmentation, the model generates 20 class scores for each image pixel, and the class with the largest score is this pixel's organ class. During the training process, a loss function (also referred to as the cost function) is needed to calculate the difference between the model classification and the real classification (ground truth). Hence, the model parameters are varied to minimize the loss function.

The softmax activation function and the cross-entropy function are applied for our V-Net model. Since we have defined 20 organclasses, for each image pixel, the model would generate a score vector Z ($Z=Z_1,..., Z_{20}$), which contains classification scores for each class. The softmax function, as shown in Equation 1, will normalize the score values ($Z_i$) into probabilities ($f(Z)_i$) which are proportional to the exponential of the scores.

$$f(Z)_i = \frac{e^{Z_i}}{\sum_{j=1}^{20} e^{Z_j}}, i = 1, \dots, 20 \qquad (1)$$

After using the softmax function to normalize the classification scores into probabilities, we further used the cross-entropy function (Equation 2) to calculate the cross-entropy between the prediction and the ground truth. The *I* represents the class, $f(Z)_i$ and $t_i$ represent predicted class probabilities and true class probabilities, and the sum ranges over the total number of classes. This cross-entropy value (CE) is the loss function of our model to evaluate the distance between prediction and ground truth.

$$CE = -\sum_{i}^{20} t_i \log(f(Z)_i) \qquad (2)$$

*2.5. Model performance metrics: Sørensen–Dice coefficient and Hausdorff Distance*

We used the Sørensen–Dice coefficient [37] [38], hereafter referred to as the Dice coefficient, as a metric to quantitatively evaluate the segmentation results of our V-Net model. The Dice coefficient (Fig. 5) is a frequently used metric to assess the similarity between two data sets. If two data sets overlap perfectly, the Dice coefficient has the maximum value of 1. Otherwise, for two independent data sets, the coefficient is zero. Therefore, a Dice coefficient close to one between the ground truth and the segmentation prediction volumes is desirable and indicates a good model performance.

Besides the Dice coefficient, we also evaluated the segmentation with a metrics based on the spatial distance between two data sets, the Hausdorff Distance (HD) (Equation 3 [39]). The HD between data set A and B is the Hausdorff distance and ‖ a-b‖ is the distance between two points. The HD value indicates the spatial distance between two data sets and lower HD values between the segmentation results and the ground truths represent better segmentation results.

$$HD(A,B) = \max(h(A,B), h(B,A))$$

$$h(A,B) = max_{(a \in A)} min_{(b \in B)} ||a - b|| \qquad (3)$$

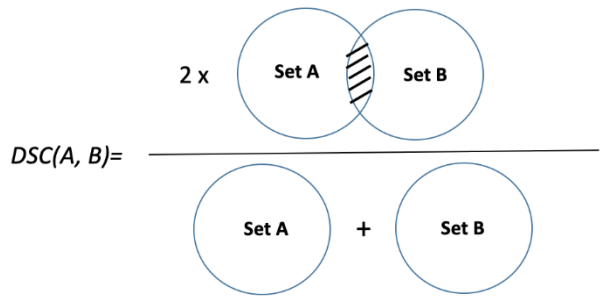

Figure 5: Schematic representation of the Sørensen–Dice coefficient for model performance evaluation.

*2.6. Improvement strategies for small-volume segmentation*

We applied several strategies to improve the segmentation accuracy of the small volumes, i.e., the lens of the eye in this work. They include enhancing the image contrast before training the model, optimizing the classification parameters in the V-Net segmentation model, and applying Mask R-CNN for the automated definition of the organ boundaries before segmentation.

*2.6.1. Deterministic image pre-processing for image quality improvement*

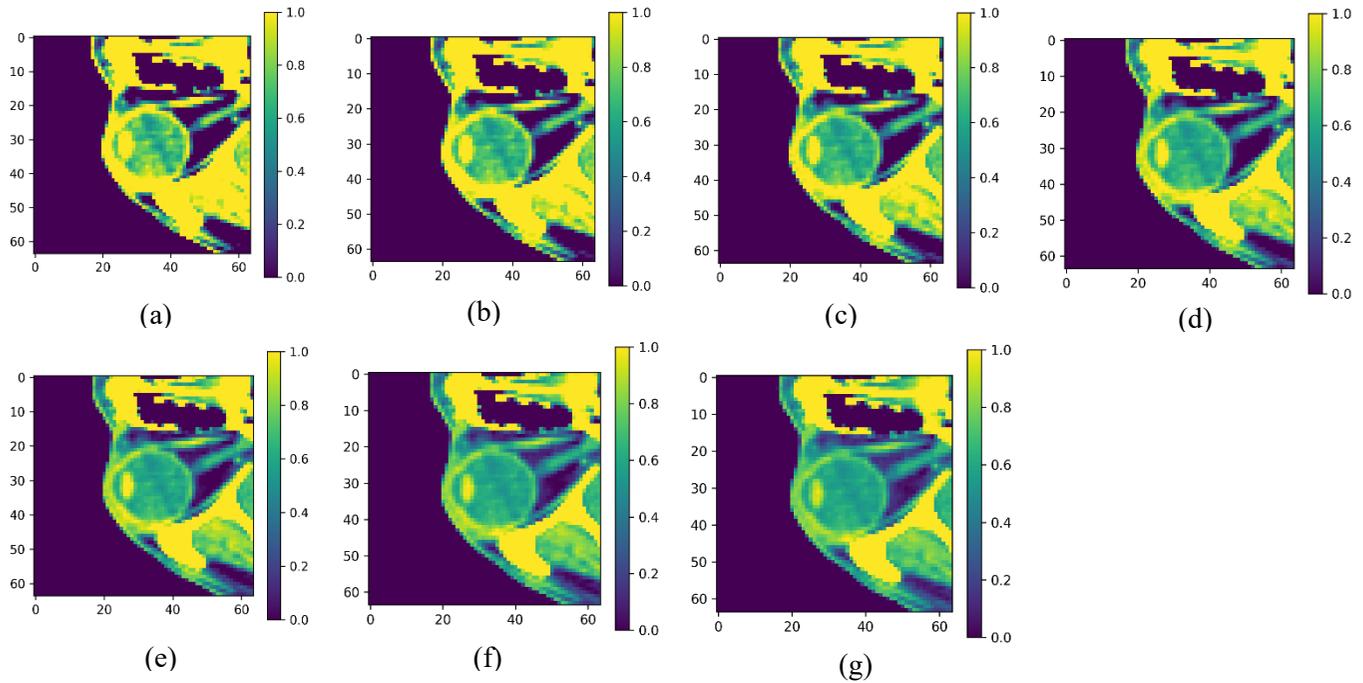

Figure 6: Eye region under different normalization ranges, (a) -40 to 40 range, (b) -50 to 50 range, (c) -60 to 60 range, (d) -70 to 70 range, (e) -80 to 80 range, (f) -90 to 90 range, (g) -100 to 100 range, the colorbar range from 0 to 1.

Due to the local variability of features of the CT scan images, data preprocessing was needed before the training to optimize the small volumes of interest and ultimately improve segmentation result. As described in Section 2, the image needs to be normalized before being analyzed by the network. Fig. 6 shows that the choice of the normalization range could significantly affect the rendering quality of the lens of the eye region. Consequently, the segmentation model would perform differently by using images with different normalization settings as the training data. Therefore, finding the optimal normalization range is expected to improve the image quality of the training data and the V-Net performance on small volumes. In this paper, we varied the normalization range from -40 to 40, to -100 to 100, in 10-level intervals to find the optimized setting.

*2.6.2. Optimization of the V-Net classification threshold*

Besides the methods described above to improve the image qualify for training, we have also tested another technique consisting of optimizing the V-Net processing parameters. The classification threshold is one parameter that could significantly affect the segmentation results. When the model segments images, it assigns classification probabilities to each pixel by the Softmax function. This value represents the probability that the pixel belongs to a certain organ. Therefore, by applying a threshold, the model will classify a pixel as a certain organ only when it has a probability larger than the threshold. If the classification probability is smaller than the threshold, the pixel is classified as background. The choice of this threshold for the lens of the eye is crucial and could significantly affect the segmentation result.

*2.6.3. Application of Mask R-CNN for the automated definition of the organ segmentation boundaries*

The third method to improve small area segmentation consisted in using a use Mask R-CNN to obtain refined boundaries of the eye regions for the segmentation. When we performed the segmentation of eyes and lens of the eyes, we first trained the model with cropped images(Fig. 3) and then used the trained weights to perform the segmentation of the corresponding cropped region in the test data set. The cropping boundaries were manually selected and kept constant across the data set. The resulting cropped region is much larger than the outline of the eye. With such a uniform cropping boundary, the organ of interest (eyes and lens of the eye) will always be included,even for slightly different eye locations. However, the potential downside of this setting is the inter-variability of structures within the cropped region so that the network cannot entirely focus onthe eyes and lens of the eye.

The method we proposed is to obtain image boundaries specific to each dataset and use these cropped eye images to train the V-Net model and perform segmentation. Since the refined eye locations are different for various patients, we have chosen to use the Mask R-CNN network to obtain the bounding box of the eyes as the cropping boundaries for each dataset. Mask R-CNN network is an object-instance segmentation network based on Faster R-CNN [33]. Faster R-CNN network can detect objects, defined"classification", and find a bounding box of each object, defined "regression". Improving upon Faster R-CNN, Mask R-CNN can alsopredict a mask of the outlines of the objects for each region of interest (RoI).

The Mask R-CNN model we applied is the Matterport Mask R-CNN [40]. We have used the head and neck image sets to train this model and used it to predict the bounding box for the eyes as the cropping boundaries. Finally, we used the V-Net model to segment the lens of the eye, specifically inside the cropping boundaries.

## 3. Results

### 3.1. V-Net segmentation results

We have developed a DL segmentation model based on V-Net to automatically segment radiological images for treatment planning. The neural network wastrained with unified settings for all organs (image pixels were normalized within the -100 to 100 range, the classification threshold was 0.8). The model was trained for 800 epochs, with 35 steps per epoch. The obtained loss value graph is shown in Fig. 7. The lossgraph reached a stable level after 800 epochs, and the model had an adequate fitting.

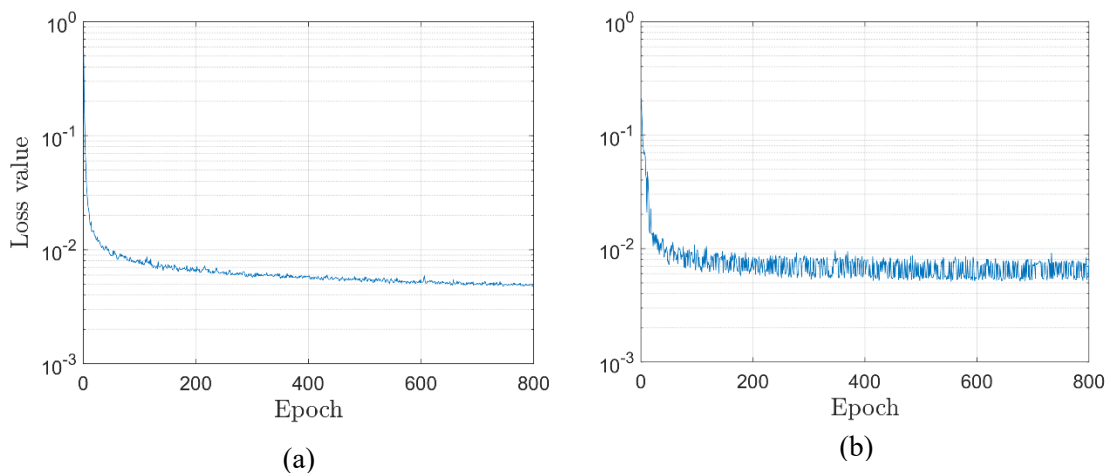

Figure 7: (a) Training loss graph and (b) validation loss graph of the V-Net segmentation model for the 800-epoch training.

After the training, our model segmented most organs well, whose prediction results were very close to the ground truth. A comparison example between the predicted segmentation and the ground truth is shown in Fig. 8. Fig. 9 shows the ground truth and the model prediction in transparency superimposed to the eye image, and we can observe an expected bias between the ground truth and the prediction for the lens of the eye.

Furthermore, Figure 10 shows the average Dice coefficients of the 10 test data sets. The corresponding region index and its region name is shown in Table 1. In Figure 10, we compared our resultsbefore the application of the small volume segmentation strategies with the results presented by W. Lei et al [23], which also used theStructSeg2019 head and neck images. The error bar is the sample standard deviation of test data sets. We can observe that most of the Dice segmentation coefficients obtained with the first version of our algorithm exhibited a high Dice coefficient hence a satisfactory segmentation and are within the sample standard deviation of the Dice coefficients reported by W. Lei, et al, except for the lens of the eye regions. The larger sample standard deviation associated with our Dice coefficients, compared to the results by W. Lei, et al, is likely due to the size of the training data set, which encompassed 35 training data sets, instead of 40, as in W. Lei et al. [23]. The relatively low Dice coefficients of the lens of the eye motivated us to apply several strategies to improve the segmentation accuracy of these small volumes, which are described in the following section. The Dice coefficient of the optical chiasma is also lower than most organs, probably due to the "X" shape of the optic chiasm.

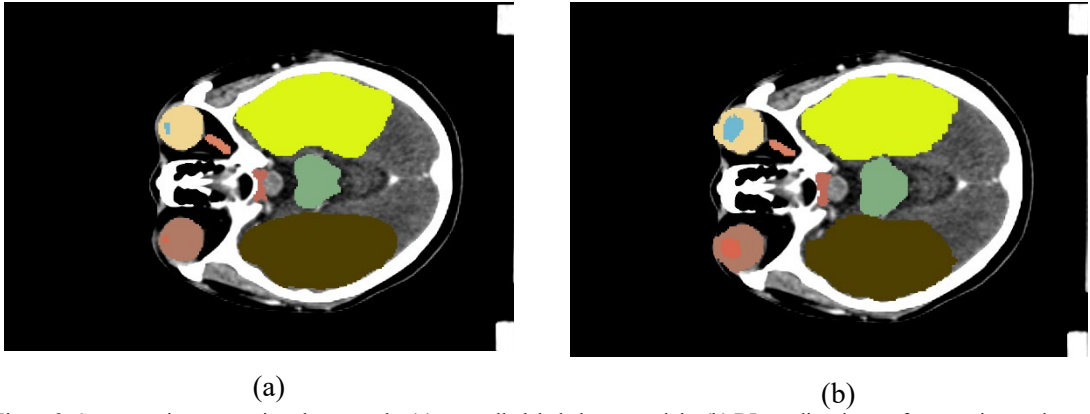

Figure 8: Segmentation comparison between the (a) manually labeled areas and the (b) DL predicted areas for a test image data.

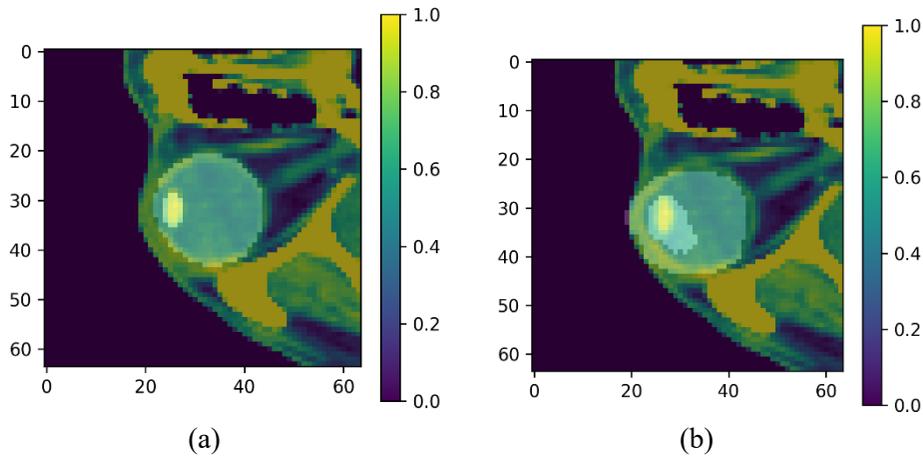

Figure 9: Segmentation comparison between the (a) manually labeled area and the (b) DL-predicted area in transparency superimposed to the eye region.

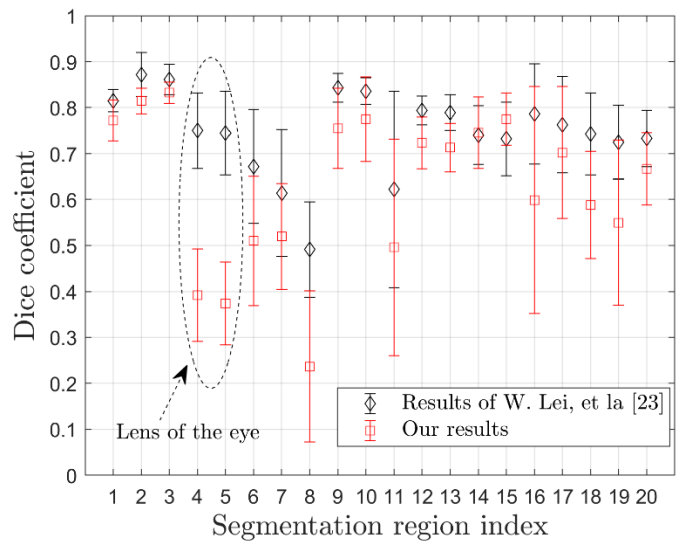

Figure 10: Average Sørensen–Dice coefficient of 20 segmented organs in the 3-D images.

Table 1: Index of the segmented soft-tissue organ and the corresponding definition.

| Index | 1 | 2 |
|---|---|---|
| Name | Brain stem | Left eye |
| Index | 3 | 4 |
| Name | Right eye | Left lens of the eye |
| Index | 5 | 6 |
| Name | Right lens of the eye | Left optic nerve |
| Index | 7 | 8 |
| Name | Right optic nerve | Optic chiasma |
| Index | 9 | 10 |
| Name | Left temporal lobes | Right temporal lobes |
| Index | 11 | 12 |
| Name | Pituitary gland | Left parotid gland |
| Index | 13 | 14 |
| Name | Right parotid gland | Left inner ear |
| Index | 15 | 16 |
| Name | Right inner ear | Left mid ear |
| Index | 17 | 18 |
| Name | Right mid Ear | Left temporomandibular joint |
| Index | 19 | 20 |
| Name | Right temporomandibular joint | Spinal cord |

*3.2. Results of the optimization of the normalization range and the classification threshold*

We have varied the image normalization range, as introduced in Section 2.6.1, to improve the segmentation performance. We normalized the pixel values of the eye region within different ranges, from -40 to 40, to -100 to 100, in 10-level intervals and kept the lens of the eye classification threshold to 0.8.

For each normalization range, we have trained the model for 800 epochs and used the seven trained models to segment the eye and the lens of the eye. A selection of the segmentation results is shown in transparency superimposed to the eye image in

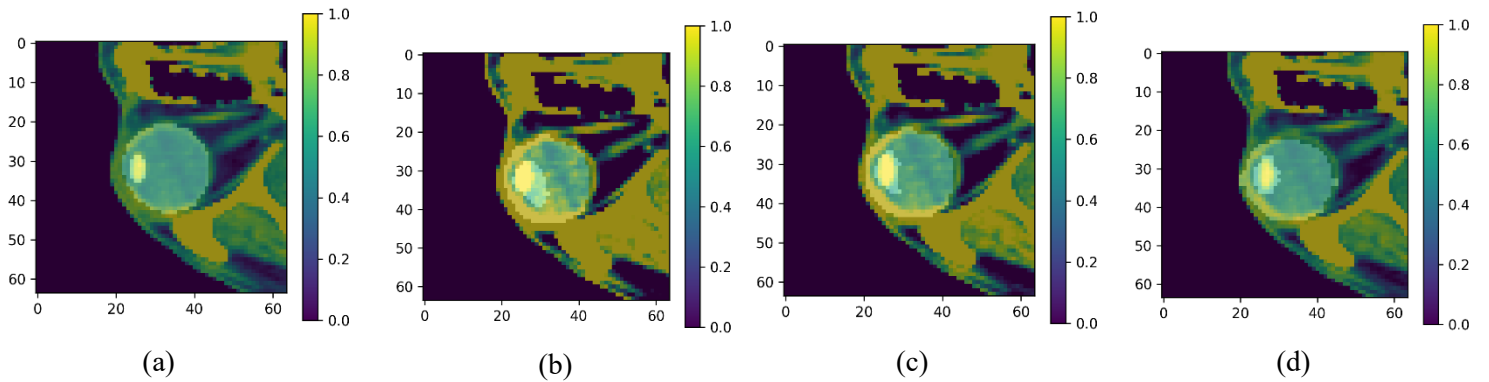

Figure 11: Segmentation results of left lens of the eye in transparency superimposed to the eye image under different normalization ranges, (a) ground truth, (b) -40 to 40 range, (c) -60 to 60 range, (d) -90 to 90 range.

Fig. 11,and it demonstrates that the normalization range settings can significantly affect the lens of the eye segmentation.

Fig. 12 shows the segmentation results under different classification threshold settings. From this prediction comparison, we can clearly observe that with a 0.75 classification threshold, more pixels are classified as the lens of the eye, and eye pixels surrounding the lenshave a higher chance to be misclassified as the lens of the eye (false positive). With a higher threshold (0.95), fewer pixels are classified as the lensof the eye, and some true lens pixels with low classification probabilities could be rejected (false negative).

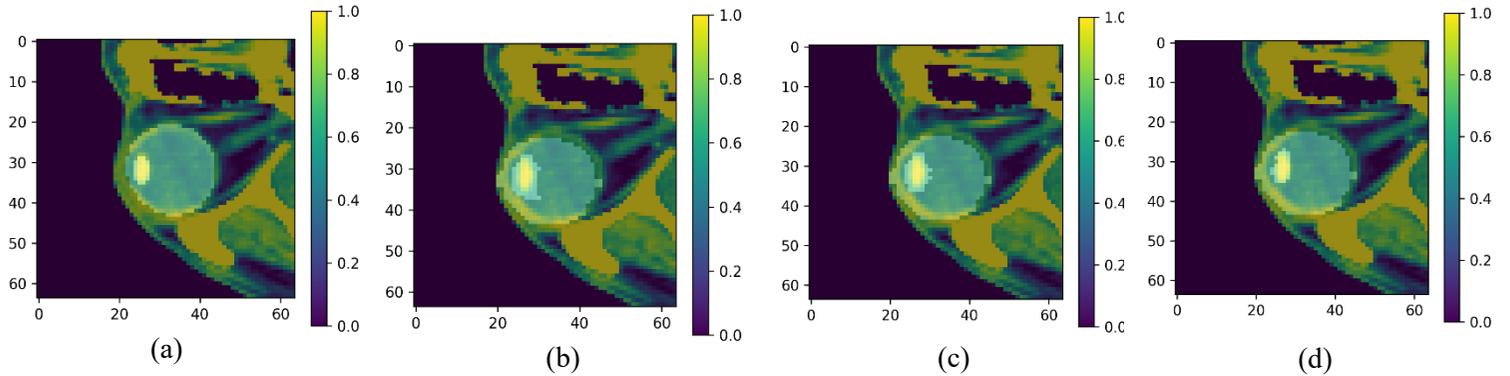

(a)　　　　　　　　　(b)　　　　　　　　　(c)　　　　　　　　　(d)

Figure 12: Left lens of the eye segmentation in transparency superimposed to the eye image under different classification thresholds, (a) ground truth, (b) 0.75 threshold,(c) 0.85 threshold, (d) 0.95 threshold, same (-90, 90) normalization range.

In order to find the optimized threshold and normalization range, we have obtained the average Dice coefficients and HD values as a bi-variate function of the settings. In Fig. 13, we found that a normalization range in the -90 to 90 range and a 0.85 classification threshold resulted in the highest Dice coefficient and lowest HD values, which represented the best result for the segmentation of the lens of the eye. With the optimized normalization range and threshold settings, the Dice coefficients of the left and right lens of the eye increased from 0.39±0.10 and 0.37±0.09 to 0.61±0.07 and 0.58±0.10 (Table 2), which correspond to a 56% improvement. The HD values decreased from 5.1±1.6 mm and 5.0±1.4 mm to 2.6±0.8 mm and 2.4±0.7 mm (Table 2), corresponding to a 50%improvement.

Table 2: Segmentation results of V-Net model for StructSeg2019 data under different improvement strategies .

| Improvement strategy | Left lens of the eye Dice coefficient (mean±std) | HD value (mean±std mm) | Right lens of the eye Dice coefficient (mean±std) | HD value (mean±std mm) |
|---|---|---|---|---|
| No strategy | 0.39±0.10 | 5.1±1.6 | 0.37±0.09 | 5.0±1.4 |
| Optimized threshold and normalization range | 0.61±0.07 | 2.6±0.8 | 0.58±0.10 | 2.4±0.7 |
| Organ-specific bounding box | 0.57±0.16 | 6.1±4.7 | 0.61±0.18 | 3.5±0.8 |

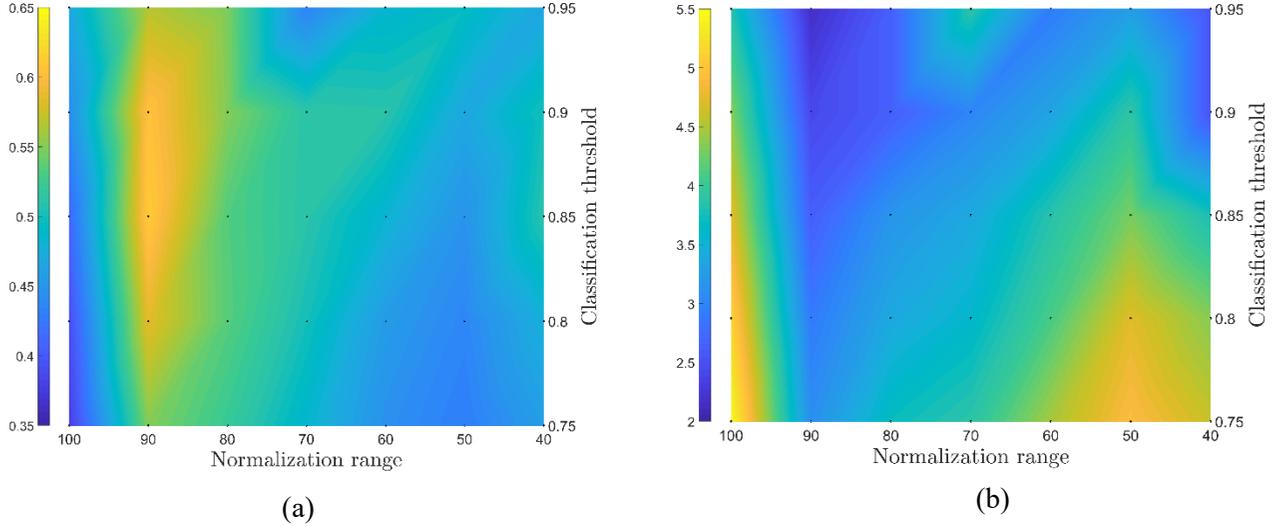

Figure 13: (a) Average Sørensen–Dice coefficients and (b) HD values of the lens of the eye for thresholds between 0.75 to 0.95 and normalization ranges between -40 to 40, to -100 to 100.

### 3.3. Eye segmentation within an organ-specific bounding box

The third method to improve the segmentation of small areas consisted in using the Mask R-CNN model to predict the bounding boxes of eyes,then using these boundaries to crop the eye regions in 3-D images, and finally, using them to train the V-Net model and perform segmentation. We processed the volumes as a sequence of 2-D slices and selected those slices that contain the eye to train the model. After 150 epochs, the validation loss of the Mask R-CNN model reached a stable value. The relative error between subsequent loss samples was lower than 10% after 150 epochs. Fig. 14(a) shows the loss graph of the training. The high initial loss value of 40 is due to the random selection of the initial weight settings. The validation loss is shown in Fig. 14(b). After 150 epochs,both loss values converged to comparable and stable values, confirming a proper model training. In some cases, the trained Mask R-CNN model predicts more than one bounding box for each lens of the eye in a single head andneck slice (Fig. 15 (a)). This effect is probably due to insufficient training data. In this case, we chose to use the union of all overlappingbounding boxes as the final bounding box, as shown in Fig. 15 (b). After defining the new bounding boxes, we cropped the eye regionsfrom the full 3-D head and neck images and reshaped them into 64×64×64 pixels for training. Fig. 16 (a) shows the cropped eye region with a Mask R-CNN predicted bounding box. A normalization range from -90 to 90 was used for this analysis since it yielded the best segmentation result, as detailed in Section 3.2. The bounding boxes around the eyes for the ten test data sets were obtained using Mask R-CNN. After applying the bounding boxes, the segmentation result is shown in Fig. 16 (b). The Dice coefficients of the left and right lens of the eye changed from 0.61±0.07 and 0.58±0.10 to 0.57±0.16 and 0.61±0.18 (Table 2). The HD values changed from 2.6±0.8 mm and 2.4±0.7 mm to 6.1±4.7 mm and 3.5±0.8 mm (Table 2). We observed that optimizing the classification threshold and the normalization range can improve the segmentation accuracy by approximately 50%, while the application of organ-specific small bounding boxes does not further improve the segmentation performance.

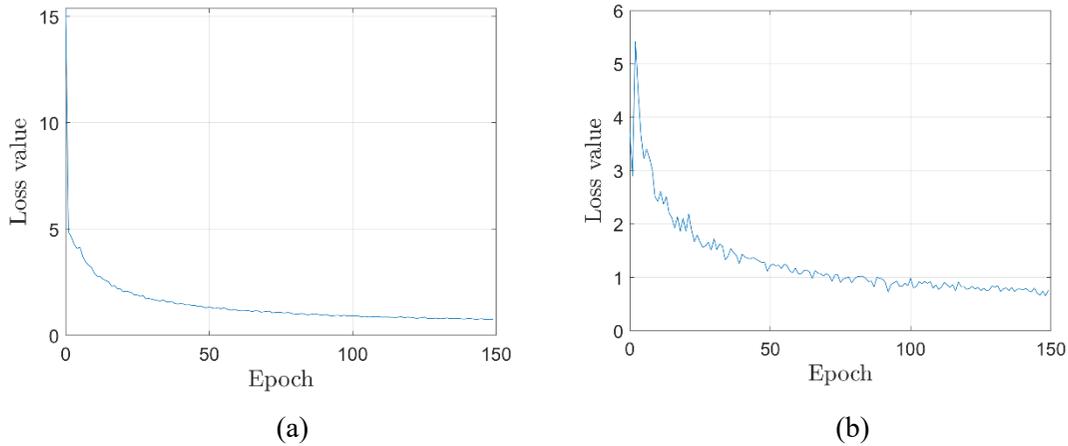

Figure 14: (a) Training loss graph and (b) validation loss graph of Mask R-CNN model for the 150-epoch training.

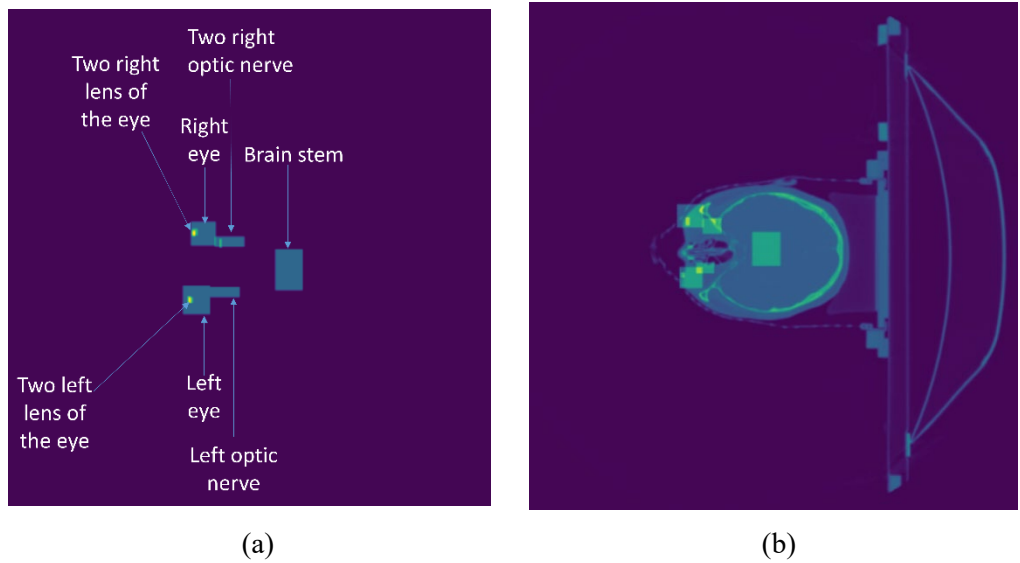

Figure 15: Bounding boxes predicted by the Mask R-CNN model for small areas in the head and neck region.

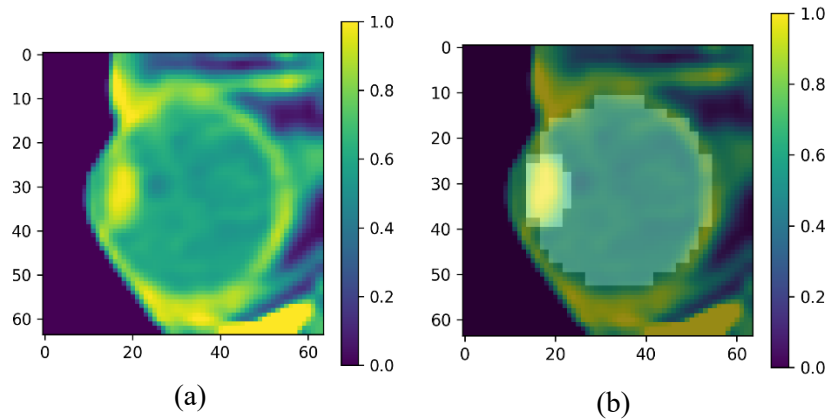

Figure 16: (a) Bounding boxes surrounding small areas in the head and neck region obtained through the Mask-R CNN network. (b) Eye region automatically cropped by the Mask R-CNN.

### 3.4. Segmentation result of external OSF healthcare images

We used the optimized model that was trained by StructSeg2019 images to segment the lens of the eye of 17 external images from OSF healthcare. We calculated the average Dice coefficients of the 17 images and show it in Fig. 17. The Dice coefficients of the left and right lens of the eye were 0.58±0.09 and 0.59±0.13, respectively, while the HD values were 3.6±1.5 mm and 3.9±1.3 mm, respectively. As the results shown, the model exhibited consistent segmentation performance for different data sets, which validates the robustness of the segmentation. After adopting the improvement strategies for small volumes, the segmentation of lens of the eye achieved satisfactory results for all test images. The 17 external images were from the clinical treatment and the size of each pixel was different for every image, which ranged from 0.44mm × 0.44mm to 0.98mm × 0.98mm. Because of the variation of the pixel size, these external test images are closer to real work clinical data than the model training data (StructSeg2019). The satisfactory results of the external images show that the robustness of the model is able to withstand variations in real clinical practice.

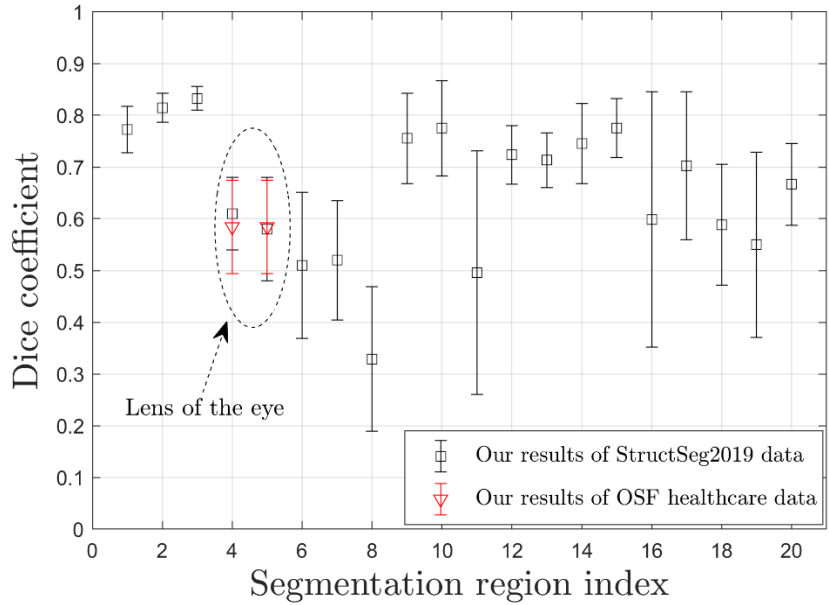

Figure 17: Segmentation results of StructSeg2019 data and OSF healthcare data.

We also studied the clinical impact of our DL segmentation method by calculating the dose received by the lens of the eye in the brain treatments and compared it to the dose delivered to the lens of the eye when using manually-contoured volumes. For the 17 OSF healthcare patients, we developed the treatment plans with parallel opposed lateral X-ray beams and blocked the bilateral lens with multi-leaf collimators. The treatment plans were calculated in Eclipse v 11.0 (Varian Medical System, Palo Alto, CA, USA). We calculated the maximum and average doses to the lens of the eye regions that were manually contoured by clinical research assistants (ground truth) and automatically contoured by the DL model (prediction). The maximum doses using the ground truth and the predicted volumes are 8.8±2.8 Gy and 13.8±6.5 Gy, respectively. The average doses to the lens of the eye are 5.2±0.9 Gy and 6.0±1.5 Gy, for manual and DL segmentation, respectively. The uncertainty is calculated as the sample standard deviation of the data set, and therefore it represents the intra-patient variability. As the results show, a treatment plan created with AI-contoured lens of the eye would result in a dose comparable to the dose delivered in case of manual contouring. This result proves the feasibility of this method for application in real clinical practice.

The computation time of our DL method was measured on Google Colab High-Ram GPUs, and the GPU in Colab includes Nvidia K80s, T4s, P4s, and P100, whose availability varies over time. The model training took approximately 12 hours, and the testing took approximately 2 minutes to segment all organs in the head and neck images and approximately 30 seconds to only segment eyes and lens of the eyes. The model training times for small and large cropped areas with various rotation angles are comparable. After proper optimization of theclassification threshold and image normalization range, their implementation negligibly affected the computation time.

**Conclusions**

In this paper, we have developed a V-Net-based model for fully-automated 3D CT image segmentation and used three methodsto improve the segmentation performance for small areas, namely the lens of the eyes in this case. We showed that the image normalization range could significantly change the shape of the segmented lens of the eye in the image. In the analyzed data set, the normalization range that resulted in the best segmentation results ranges from -90 to 90 intensity levels. Besides the normalization settings, the classification threshold settings can also significantly affect the lens of the eye segmentation. With an optimized threshold setting of 0.85, the model could reject as many misclassified image pixels as possible while retaining most of the pixels in the lens of the eye in the ground truth labels, manually segmented by expert radiologists. We have then used the Mask R-CNN algorithm to automatically define tight bounding boxes surrounding the eyes and then used such boundaries to crop the V-Net training images. The result indicates that the application of the small boundaries cannot further improve the segmentation accuracy of small volumes.

By applying the aforementioned methods, we achieved a significant improvement of the Dice coefficient of the lens of the eye from 0.39±0.10 and 0.37±0.90 to 0.61±0.70 and 0.58±0.10 and the HD values from 5.1±1.6 mm and 5.0±1.4 mm to 2.6±0.8 mm and 2.4±0.7 mm for the left and right lens of the eye, respectively. While the specific parameters may not generally be applicable to all organs, we demonstrated three specific strategies, combined with a V-Net, that allow improving the segmentation of small areas. The robustness of the model for small volume segmentation was further confirmed through segmenting real-world clinical data acquired with different hardware and protocols compared to the training data set. Besides evaluating the model performance with metrics like the Dice coefficient, we also calculated the dose received by the lens of the eye in brain treatments to study the clinical impact of the AI contouring. The agreement of the dose to the lens between manual and DL segmentation and the consistent segmentation performance for images under various protocols and demonstrated the feasibility of using the DL algorithm in clinic practice. Improvement of small area segmentation and hence the application of our approaches can be important in the treatment planning of target volumes in close proximity to small radio-sensitive regions or in pediatric patients. Semi-automated model-based [41] and atlas-based [42] semi-automatic segmentation methods achieve Dice coefficients of approximately 0.7 for the lens of the eye, which outperform our deep-learning based model without any improvement strategies for small areas. After the improvement, our model achieved comparable segmentation accuracy of the lens of the eye, with respect to semi-automated methods. Therefore, we demonstrated that specific segmentation strategies can be implemented in fully-automated deep-learning-based segmentation, while retaining the superior robustness of the DL based approach, compared to semi-automated and manual methods [43].

Even though the achieved Dice coefficient of the lens of the eye is lower compared to other head and neck regions, we have presented a systematic approach to study the sensitivity of the segmentation to multiple parameters and demonstrated that improvement in the small volumes could be achieved by carefully optimizing parameters such as the image intensity levels and the algorithm classification threshold. The proposed methods can be generalized to other regions and are independent of the imaging modality. From a computational perspective, the proposed methods do not require a significantly longer processing time compared to the initial V-Net segmentation.

In current treatment planning protocols, target volumes are typically contoured on CT images before the treatment manually or using semi-automated software. A margin of a few mm between the clinical target volume and the planning target volume is applied to account for potential target volume and shape changes during the treatment fractions. This approach unavoidably increased the dose to healthy tissue surrounding the treatment volume while not maximizing the dose to the tumor. Fully validated deep-learning segmentation could enable patient-specific adaptive daily segmentation to minimize the risks associated with over treatment of the planning target volume. With the decreased PTV and toxicity, we can improve the accuracy and precision of target hitting, and therefore, significantly improve the tumor control in the current practice of cancer treatment.

**Acknowledgement**
This work was funded in part by the Nuclear Regulatory Commission Faculty Development Grant number 31310019M0011 and the Jump ARCHES endowment through the Health Care Engineering Systems Center.

**References**

[1] W. D. Newhauser and M. Durante, "Assessing the risk of second malignancies after modern radiotherapy," 2011.

[2] A. Di Fulvio, L. Tana, M. Caresana, E. D'Agostino, M. De San Pedro, C. Domingo, and F. D'Errico, "Clinical simulations of prostate radiotherapy using BOMAB-like phantoms: Results for neutrons," *Radiation Measurements*, 2013.

[3] S. Miljanić, I. Bessieres, J. M. Bordy, F. D'Errico, A. Di Fulvio, D. Kabat, Z. Knežević, P. Olko, L. Stolarczyk, L. Tana, and R. Harrison, "Clinical simulations of prostate radiotherapy using BOMAB-like phantoms: Results for photons," *RadiationMeasurements*, 2013.

[4] J. M. Bordy, I. Bessieres, E. D'Agostino, C. Domingo, F. D'Errico, A. Di Fulvio, Ž. Knežević, S. Miljanić, P. Olko, A. Os- trowsky, B. Poumarede, S. Sorel, L. Stolarczyk, and D. Vermesse, "Radiotherapy out-of-field dosimetry: Experimental and computational results for photons in a water tank," *Radiation Measurements*, 2013.

[5] W. Fu, Y. Yang, N. J. Yue, D. E. Heron, and M. S. Huq, "A cone beam CT-guided online plan modification technique to correct interfractional anatomic changes for prostate cancer IMRT treatment," *Physics in Medicine and Biology*, 2009.

[6] G. P. Chen, G. Noid, A. Tai, F. Liu, C. Lawton, B. Erickson, and X. A. Li, "Improving CT quality with optimized image parameters for radiation treatment planning and delivery guidance," *Physics and Imaging in Radiation Oncology*, 2017.

[7] F. X. Li, J. B. Li, Y. J. Zhang, T. H. Liu, S. Y. Tian, M. Xu, D. P. Shang, and C. S. Ma, "Comparison of the planning target volume based on three-dimensional CT and four-dimensional CT images of non-small-cell lung cancer," *Radiotherapy and Oncology*, 2011.

[8] A. Tzikas, P. Karaiskos, N. Papanikolaou, P. Sandilos, E. Koutsouveli, E. Lavdas, C. Scarleas, K. Dardoufas, B. K. Lind, and
P. Mavroidis, "Investigating the clinical aspects of using CT vs. CT-MRI images during organ delineation and treatment planningin prostate cancer radiotherapy," *Technology in Cancer Research and Treatment*, 2011.

[9] E. Rietzel, G. T. Chen, N. C. Choi, and C. G. Willet, "Four-dimensional image-based treatment planning: Target volume segmentation and dose calculation in the presence of respiratory motion," *International Journal of Radiation Oncology BiologyPhysics*, 2005.

[10] L. Wang, S. Hayes, K. Paskalev, L. Jin, M. K. Buyyounouski, C. C. Ma, and S. Feigenberg, "Dosimetric comparison of stereo-tactic body radiotherapy using 4D CT and multiphase CT images for treatment planning of lung cancer: Evaluation of the impacton daily dose coverage," *Radiotherapy and Oncology*, 2009.

[11] V. Grégoire, P. Levendag, K. K. Ang, J. Bernier, M. Braaksma, V. Budach, C. Chao, E. Coche, J. S. Cooper, G. Cosnard, A. Eis- bruch, S. El-Sayed, B. Emami, C. Grau, M. Hamoir, N. Lee, P. Maingon, K. Muller, and H. Reychler, "CT-based delineation of lymph node levels and related CTVs in the node-negative neck: DAHANCA, EORTC, GORTEC, NCIC, RTOG consensusguidelines," *Radiotherapy and Oncology*, 2003.

[12] V. Grégoire, A. Eisbruch, M. Hamoir, and P. Levendag, "Proposal for the delineation of the nodal CTV in the node-positive andthe post-operative neck," *Radiotherapy and Oncology*, 2006.

[13] T. Heimann and H. P. Meinzer, "Statistical shape models for 3D medical image segmentation: A review," *Medical Image

*Analysis*, 2009.

[14] J. J. Cerrolaza, M. Reyes, R. M. Summers, M. Á. González-Ballester, and M. G. Linguraru, "Automatic multi-resolution shapemodeling of multi-organ structures," *Medical Image Analysis*, 2015.

[15] A. R. Eldesoky, E. S. Yates, T. B. Nyeng, M. S. Thomsen, H. M. Nielsen, P. Poortmans, C. Kirkove, M. Krause, C. Kamby, I. Mjaaland, E. S. Blix, I. Jensen, M. Berg, E. L. Lorenzen, Z. Taheri-Kadkhoda, and B. V. Offersen, "Internal and external validation of an ESTRO delineation guideline – dependent automated segmentation tool for loco-regional radiation therapy of early breast cancer," *Radiotherapy and Oncology*, 2016.

[16] J. F. Daisne and A. Blumhofer, "Atlas-based automatic segmentation of head and neck organs at risk and nodal target volumes:A clinical validation," *Radiation Oncology*, 2013.

[17] P. Aljabar, R. A. Heckemann, A. Hammers, J. V. Hajnal, and D. Rueckert, "Multi-atlas based segmentation of brain images: Atlas selection and its effect on accuracy," *NeuroImage*, 2009.

[18] D. Rueckert and J. A. Schnabel, "Registration and segmentation in medical imaging," *Studies in Computational Intelligence*, 2014.

[19] J. Long, E. Shelhamer, and T. Darrell, "Fully convolutional networks for semantic segmentation," in *Proceedings of the IEEE Computer Society Conference on Computer Vision and Pattern Recognition*, 2015.

[20] O. Ronneberger, P. Fischer, and T. Brox, "U-net: Convolutional networks for biomedical image segmentation," in *Lecture Notes in Computer Science (including subseries Lecture Notes in Artificial Intelligence and Lecture Notes in Bioinformatics)*, 2015.

[21] E. Gibson, F. Giganti, Y. Hu, E. Bonmati, S. Bandula, K. Gurusamy, B. Davidson, S. P. Pereira, M. J. Clarkson, and D. C. Barratt, "Automatic Multi-Organ Segmentation on Abdominal CT with Dense V-Networks," *IEEE Transactions on Medical Imaging*, 2018.

[22] P. H. Conze, A. E. Kavur, E. Cornec-Le Gall, N. S. Gezer, Y. Le Meur, M. A. Selver, and F. Rousseau, "Abdominal multi-organ segmentation with cascaded convolutional and adversarial deep networks," *Artificial Intelligence in Medicine*, vol. 117, 2021.

[23] W. Lei, H. Mei, Z. Sun, S. Ye, R. Gu, H. Wang, R. Huang, S. Zhang, S. Zhang, and G. Wang, "Automatic segmentation of organs- at-risk from head-and-neck CT using separable convolutional neural network with hard-region-weighted loss," *Neurocomputing*, vol. 442, pp. 184–199, 2021.

[24] W. Zhu, Y. Huang, L. Zeng, X. Chen, Y. Liu, Z. Qian, N. Du, W. Fan, and X. Xie, "AnatomyNet: Deep learning for fast and fully automated whole-volume segmentation of head and neck anatomy," *Medical Physics*, 2019.

[25] Y. Zhong, Y. Yang, Y. Fang, J. Wang, and W. Hu, "A Preliminary Experience of Implementing Deep-Learning Based Auto- Segmentation in Head and Neck Cancer: A Study on Real-World Clinical Cases," *Frontiers in Oncology*, vol. 11, 2021.

[26] E. Tappeiner, S. Pröll, M. Hönig, P. F. Raudaschl, P. Zaffino, M. F. Spadea, G. C. Sharp, R. Schubert, and K. Fritscher,

"Multi-organ segmentation of the head and neck area: an efficient hierarchical neural networks approach," *International Journal of Computer Assisted Radiology and Surgery*, vol. 14, no. 5, pp. 745–754, 2019.

[27] Y. Gao, R. Huang, M. Chen, Z. Wang, J. Deng, Y. Chen, Y. Yang, J. Zhang, C. Tao, and H. Li, "FocusNet: Imbalanced Largeand Small Organ Segmentation with an End-to-End Deep Neural Network for Head and Neck CT Images," in *Lecture Notes in Computer Science (including subseries Lecture Notes in Artificial Intelligence and Lecture Notes in Bioinformatics)*, vol. 11766LNCS, pp. 829–838, 2019.

[28] H. Seo, C. Huang, M. Bassenne, R. Xiao, and L. Xing, "Modified U-Net (mU-Net) with Incorporation of Object-Dependent High Level Features for Improved Liver and Liver-Tumor Segmentation in CT Images," *IEEE Transactions on Medical Imaging*, vol. 39, no. 5, pp. 1316–1325, 2020.

[29] B. V. Worgul, Y. I. Kundiyev, N. M. Sergiyenko, V. V. Chumak, P. M. Vitte, C. Medvedovsky, E. V. Bakhanova, A. K. Junk,
O. Y. Kyrychenko, N. V. Musijachenko, S. A. Shylo, O. P. Vitte, S. Xu, X. Xue, and R. E. Shore, "Cataracts among Chernobylclean-up workers: Implications regarding permissible eye exposures," *Radiation Research*, vol. 167, no. 2, pp. 233–243, 2007.

[30] G. Chodick, N. Bekiroglu, M. Hauptmann, B. H. Alexander, D. M. Freedman, M. M. Doody, L. C. Cheung, S. L. Simon, R. M.Weinstock, A. Bouville, and A. J. Sigurdson, "Risk of cataract after exposure to low doses of ionizing radiation: A 20-year prospective cohort study among US radiologic technologists," *American Journal of Epidemiology*, vol. 168, no. 6, pp. 620–631,2008.

[31] I. Bruchmann, B. Szermerski, R. Behrens, and L. Geworski, "Impact of radiation protection means on the dose to the lens of theeye while handling radionuclides in nuclear medicine," *Zeitschrift für Medizinische Physik*, vol. 26, no. 4, pp. 298–303, 2016.

[32] F. Milletari, N. Navab, and S. A. Ahmadi, "V-Net: Fully convolutional neural networks for volumetric medical image segmen-tation," *Proceedings - 2016 4th International Conference on 3D Vision, 3DV 2016*, pp. 565–571, 2016.

[33] K. He, G. Gkioxari, P. Dollár, and R. Girshick, "Mask R-CNN," *IEEE Transactions on Pattern Analysis and Machine Intelli- gence*, vol. 42, no. 2, pp. 386–397, 2020.

[34] "Automatic structure segmentation for radiotherapy planning challenge 2019!." https://structseg2019.grand-challenge.org/, 2019.

[35] J.~S.~Bridle, "Training stochastic model recognition algorithms as networks can lead to maximum mutual information estimationof parameters," in *Advances in Neural Information Processing Systems*, 1990.

[36] D. M. Hutton, "The Cross-Entropy Method: A Unified Approach to Combinatorial Optimisation, Monte-Carlo Simulation andMachine Learning," 2005.

[37] T. Sørensen, "A method of establishing groups of equal amplitude in plant sociology based on similarity of species content, and its application to analyses of the vegetation on Danish commons," *Kongelige Danske Videnskabernes Selskabs BiologiskeSkrifter*, vol. 5, no. 4, 1948.

[38] L. R. Dice, "Measures of the Amount of Ecologic Association Between Species," *Ecology*, vol. 26, no. 3, 1945.

[39] A. A. Taha and A. Hanbury, "Metrics for evaluating 3D medical image segmentation: Analysis, selection, and tool," *BMC Medical Imaging*, vol. 15, no. 1, 2015.

[40] W. Abdulla, "Mask r-cnn for object detection and instance segmentation on keras and tensorflow." https://github.com/matterport/Mask_RCNN, 2017.

[41] M. B. Cuadra, S. Gorthi, F. I. Karahanoglu, B. Paquier, A. Pica, H. P. Do, A. Balmer, F. Munier, and J. P. Thiran, "Model-based segmentation and fusion of 3D computed tomography and 3D ultrasound of the eye for radiotherapy planning," in *Computational Methods in Applied Sciences*, vol. 19, pp. 247–263, 2011.

[42] V. Fortunati, R. F. Verhaart, F. Van Der Lijn, W. J. Niessen, J. F. Veenland, M. M. Paulides, and T. Van Walsum, "Tissue segmentation of head and neck CT images for treatment planning: A multiatlas approach combined with intensity modeling," *Medical Physics*, vol. 40, no. 7, 2013.

[43] A. E. Kavur, N. S. Gezer, M. Barış, Y. Şahin, S. Özkan, B. Baydar, U. Yüksel, Ç. Kılıkçıer, Olut, G. B. Akar, G. Ünal, O. Dicle, and M. A. Selver, "Comparison of semi-automatic and deep learning-based automatic methods for liver segmentation in livingliver transplant donors," *Diagnostic and Interventional Radiology*, vol. 26, no. 1, pp. 11–21, 2020.